\documentclass{ws-ijbc}
\usepackage{graphicx}
\usepackage{epstopdf}
\begin{document}

\catchline{}{}{}{}{} 

\markboth{Ovsyannikov et al.}{Bifurcation to chaos}

\title{Bifurcation to Chaos in the complex Ginzburg-Landau equation with large third-order dispersion.}

\author{Ivan Ovsyannikov}

\address{Institute of Applied Mathematics and Cybernetics\\
Lobachevsky State University of Nizhny Novgorod\\
ul. Ulyanova 10, Nizhny Novgorod 603005, Russia\\
ivan.i.ovsyannikov@gmail.com}

\author{Dmitry Turaev\footnote{This work was supported by grant 14-41-00044 of the RSF.}}
\address{Mathematics Department, Imperial College, London, SW7 2AZ, UK\\
dturaev@imperial.ac.uk}

\author{Sergey Zelik}
\address{Mathematics Department, University of Surrey, Guildford, GU2 7XH, UK\\
szelik@surrey.ac.uk}

\maketitle

\begin{history}
\received{(to be inserted by publisher)}
\end{history}

\begin{abstract}
We give an analytic proof of the existence of Shilnikov chaos in complex Ginzburg-Landau equation
subject to a large third-order dispersion perturbation.
\end{abstract}


The goal of this paper is to show that chaotic behavior is possible in complex Ginzburg-Landau equation
with additional large dispersion term. The equation is
\begin{equation}\label{1}
\partial_t u=(1+i\gamma)\partial^2_x u+\beta u-(1+i\omega)u|u|^2+L\partial^3_x u,
\end{equation}
where $u$ is a complex-values function, spatially periodic with period $2\pi$, i.e. we consider Eq. (\ref{1})
on the interval $x\in (-\pi, \pi)$ with the periodic boundary condition $u(-\pi)=u(\pi)$. The dispersion term
$L u_{xxx}$ causes fast temporal oscillations in the solution, so the evolution is described by effective averaged equation
(see Eq. (\ref{2})). This averaging is performed in Ref. \cite{TiZe}; it was also shown there that the averaging in a presence of
the second-order dispersion term $iL u_{xx}$ with large $L$ leads to a significant simplification of the dynamics (the averaged system acquires a gradient
structure). In this letter we show that, surprisingly, introducing the dispersion term as in Eq. (\ref{1}) does not make dynamics
gradient, and chaos can emerge in the averaged system. As the result is of an ideological nature, we refrained of the use of
numerical integration. Instead, we provide an analytic proof of the existence of chaos in the averaged system (assisted by Maple and Mathematica tools).

As $L$ is large, Eq. (\ref{1}) can be viewed as a perturbation of the auxiliary linear dispersion equation
\begin{equation}\label{ld}
u_t=Lu_{xxx}.
\end{equation}
By choosing the orthogonal basis ${\bf e}_n=e^{inx}$, $n\in Z$, the flow $\mathcal H_L(t)$ generated by Eq. (\ref{ld}) is given by
$$\mathcal H_L(t) {\bf e}_n =e^{in^3 Lt} {\bf e}_n.$$
These solutions are $2\pi/L$-periodic with respect to time, i.e. they correspond to fast oscillations.

In order to average these oscillations, one makes the following change of variable $u$ in Eq. (\ref{1}):
$$u(t) = \mathcal H_L(t)w(t)$$
The equations for $w$ acquire explicit rapidly oscillating terms. Averaging them out is done by \citet{TiZe}. The result is the following
equation:
\begin{equation}\label{2}
\partial_t w=(1+i\gamma)\partial^2_x w+\beta w-(1+i\omega) N(w),
\end{equation}
with the operator $N$ given by
$$N(w)=\left(2w \sum_{n\in Z} |w_n|^2+\bar w \sum_{n\in Z} w_n w_{-n}-2w_0 |w_0|^2\right) {\bf e}_0- \sum_{n\neq 0} w_n(|w_n|^2+2|w_{-n}|^2) {\bf e}_n,$$
where we denote $w=\sum_{n\in Z} w_n {\bf e}_n$, and $\bar w$ is the complex conjugate of $w$.

This is an infinite-dimensional system. It is shown in Ref. \cite{TiZe} that it is well-posed and has a global attractor in an appropriate Sobolev space.
The study of its full dynamics can be difficult, however this system has finite-dimensional invariant manifolds. One of these manifolds is
$$w_n=0 \qquad \mbox{for all} \qquad |n|\geq 2.$$
In restriction to this space, we have $w=y {\bf e}_0+ v({\bf e}_1+{\bf e}_{-1})$. Then
$$N(w)=(4y |v|^2+y|y|^2+2\bar y v^2) {\bf e}_0+ (v(2|y|^2+3|v|^2)+\bar v y^2)({\bf e}_1+{\bf e}_{-1}),$$
and Eq. (\ref{2}) becomes
$$\left\{\begin{array}{l}
\dot y= \beta y -(1+i\omega)\left[y(|y|^2+4|v|^2)+2\bar y v^2\right],\\
\dot v=(\beta-1-i\gamma)v-(1+i\omega)\left[v(2|y|^2+3|v|^2)+\bar v y^2\right].
\end{array}\right.$$
Let $y=\sqrt{r}e^{i\varphi}$, $v=\sqrt{\rho}e^{i\psi}$. We obtain
$$\frac{\dot r}{2\sqrt{r}}e^{i\varphi}+i\dot\varphi \sqrt{r}e^{i\varphi}
= \beta \sqrt{r}e^{i\varphi} -(1+i\omega)\left[\sqrt{r}e^{i\varphi}(r+4\rho)+
2\sqrt{r}e^{-i\varphi} \rho e^{2i\psi}\right],$$
which gives
$$\dot r= 2r\left[\beta -r-4\rho-2\rho (\cos\eta-\omega\sin\eta)\right],$$
$$\dot\varphi = -\omega(r+4\rho)-2(\sin\eta+\omega\cos\eta)\rho,$$
where $\eta=2(\psi-\varphi)$. Similarly,
$$\frac{\dot \rho}{2\sqrt{\rho}}e^{i\psi}+i\dot\psi \sqrt{\rho}e^{i\psi}
=(\beta-1-i\gamma)\sqrt{\rho}e^{i\psi}-(1+i\omega)\left[\sqrt{\rho}e^{i\psi}(2r+3\rho)+
\sqrt{\rho}e^{-i\psi} r e^{2i\varphi}\right],$$
so
$$\dot \rho=2\rho\left[\beta-1-2r-3\rho-(\cos\eta+\omega\sin\eta)r\right],$$
$$\dot\psi =-\gamma-\omega(2r+3\rho)+(\sin\eta-\omega\cos\eta)r.$$

Finally (by scaling time to $2$), we obtain the following three-dimensional system:
\begin{equation}\label{ssm}
\left\{\begin{array}{l}\dot r= r\left[\beta -r-4\rho-2\rho (\cos\eta-\omega\sin\eta)\right],\\
\dot \rho=\rho\left[\beta-1-2r-3\rho-(\cos\eta+\omega\sin\eta)r\right],\\
\dot\eta =-\gamma+\omega(\rho-r)+(\sin\eta-\omega\cos\eta)r+2(\sin\eta+\omega\cos\eta)\rho.\end{array}\right.
\end{equation}

\begin{figure}
\centerline{\includegraphics[width=10 cm]{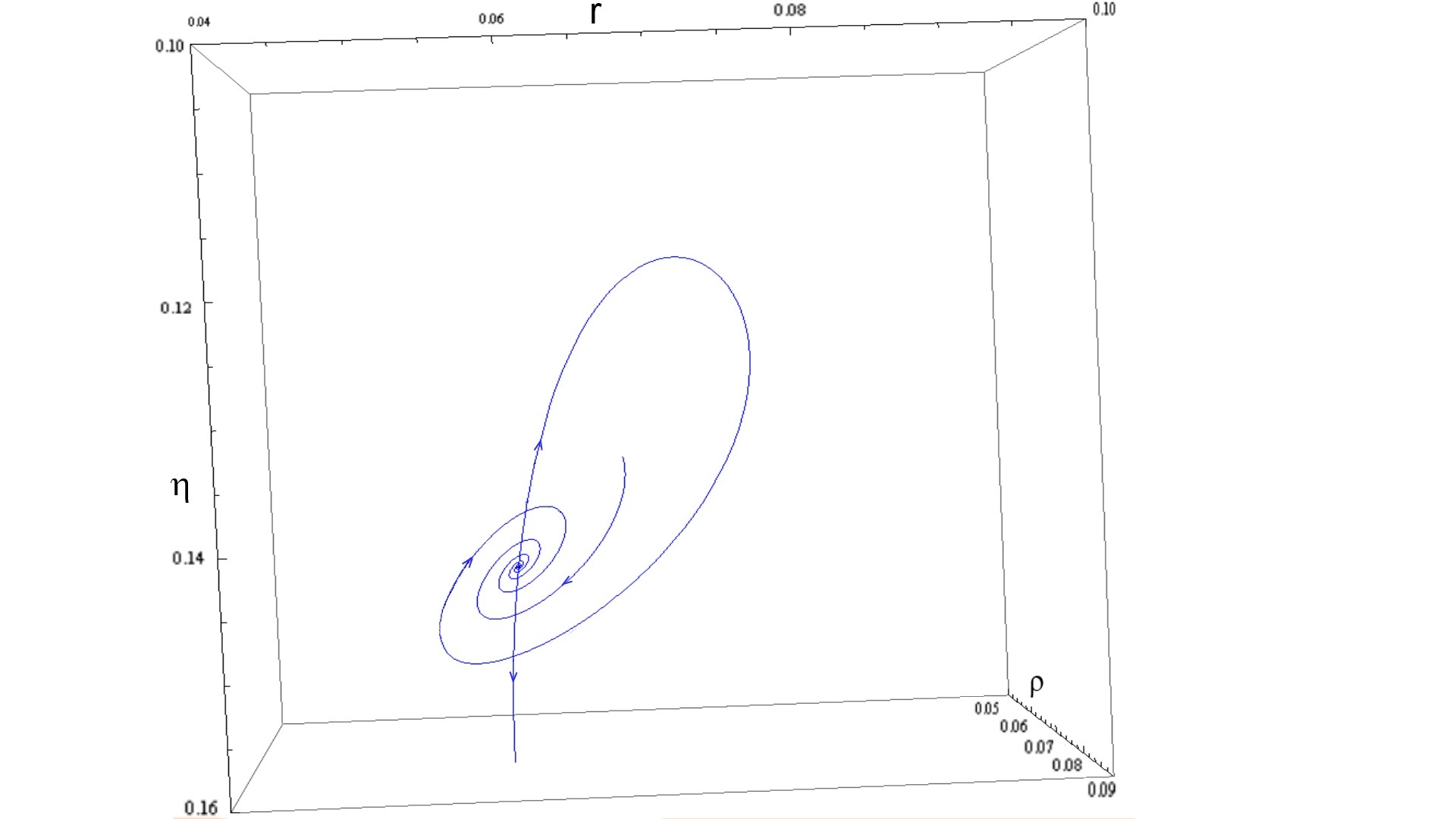}}
\caption{Homoclinic loop to the saddle-focus in system (\ref{ssm}). The equilibrium is at $r = 0.06$, $\rho = 0.07221$, $\eta = 0.143345$, which
corresponds to $\omega\approx -30.965$, $\beta \approx 1.1306$, $\gamma\approx -2.9361$ (see Eqs. (\ref{bgm}),(\ref{equ})).}
\end{figure}

The computations that follow show that system (\ref{ssm}) has a region of parameter values $(\beta,\gamma,\omega)$ which correspond
to chaotic behavior. We prove this by showing that at a certain parameter value the system has Shilnikov saddle-focus homocinic loop
\cite{Shi65,Shi70} (see also Ref. \cite{HShi}). Doing this numerically would be an easy exercise -- e.g. see in Fig. 1
a homoclinic loop we found in this system. However, we want to prove its existence analytically. To this aim, we employ the idea
of \citet{Tresser,Arneodo} who showed that bifurcations of an equilibrium state with triple zero eigenvalue can lead to the birth of
the Shlnikov loop. Fully analytic proof of this can be obtained using the result by \citet{Ibanez}; we also mention that this method
was used for an analytic proof of (space-time) chaos in Ginzburg-Landau type equations by \citet{TuZe}.

Thus, in the rest of the paper we are showing that system (\ref{ssm}) has values of parameters which correspond to the equilibrium state
that has all three eigenvalues zero and satisfies the conditions of Arneodo-Coullet-Spiegel-Tresser-Ibanez-Rodriguez theorem. It is easy to see
that at non-zero $r$ and $\rho$, the equilibria are given by
\begin{equation}\label{bgm}
\begin{array}{l}
\beta = r+4\rho+2\rho (\cos\eta-\omega\sin\eta),\\
\beta=1+2r+3\rho+(\cos\eta+\omega\sin\eta)r,\\
\gamma=\omega(\rho-r)+(\sin\eta-\omega\cos\eta)r+2(\sin\eta+\omega\cos\eta)\rho.\end{array}
\end{equation}
Thus, any $(r,\rho,\eta)$ can be an equilibrium for an appropriate choice of $\beta$ and $\gamma$, provided
\begin{equation}\label{equ}
\rho(1+2(\cos\eta-\omega\sin\eta))-r(1+(\cos\eta+\omega\sin\eta))=1.
\end{equation}

The linearization matrix at the non-zero equilibrium is
\begin{equation}\label{matr}
A=\left(\begin{array}{ccc}
-r & -2r(2+\cos\eta-\omega\sin\eta) & 2 r\rho (\sin\eta+\omega\cos\eta)\\
-\rho (2+\cos\eta+\omega\sin\eta) & -3\rho & r\rho(\sin\eta-\omega\cos\eta)\\
-\omega+\sin\eta-\omega\cos\eta & \omega+2(\sin\eta+\omega\cos\eta) & (\cos\eta+\omega\sin\eta)r+2(\cos\eta-\omega\sin\eta)\rho
\end{array}\right).
\end{equation}

We now look for the values of $r,\rho,\eta$ that correspond to a triple zero eigenvalue of $A$. This happens when
the trace, determinant, and the sum $\sigma A$ of the three main second-order minors of $A$ are simultaneously zero.

Condition $tr A=0$ reads as
\begin{equation}\label{tra}
r(1-\cos\eta-\omega\sin\eta)+\rho(3-2\cos\eta+2\omega\sin\eta)=0.
\end{equation}
Note that if $\sin\eta=0$, then $tr A$ cannot vanish at positive $r$ and $\rho$,
so we further assume $\sin\eta\neq 0$.

The determinant of matrix $A$ vanishes when
$$\rho J_1+rJ_2=0$$
where $$J_1=
\left|\begin{array}{ccc}
-1 & -2(2+\cos\eta-\omega\sin\eta) & 2 (\sin\eta+\omega\cos\eta)\\
-(2+\cos\eta+\omega\sin\eta) & -3 & 0\\
-\omega+\sin\eta-\omega\cos\eta & \omega+2(\sin\eta+\omega\cos\eta) & 2(\cos\eta-\omega\sin\eta)
\end{array}\right|,$$
and
$$J_2= \left|\begin{array}{ccc}
-1 & -2(2+\cos\eta-\omega\sin\eta) & 0\\
-(2+\cos\eta+\omega\sin\eta) & -3 & \sin\eta-\omega\cos\eta\\
-\omega+\sin\eta-\omega\cos\eta & \omega+2(\sin\eta+\omega\cos\eta) & \cos\eta+\omega\sin\eta\end{array}\right|.
$$
We have
$$J_1=-2(2+\cos\eta+\omega\sin\eta)\left|\begin{array}{cc}
4+2(\cos\eta-\omega\sin\eta) & \sin\eta+\omega\cos\eta\\
-\omega-2(\sin\eta+\omega\cos\eta) & \cos\eta-\omega\sin\eta\end{array}\right|-$$
$$\qquad\qquad-6\left|\begin{array}{cc}
-1 & \sin\eta+\omega\cos\eta\\
-\omega+\sin\eta-\omega\cos\eta & \cos\eta-\omega\sin\eta\end{array}\right|=$$
$$=-(4+2\cos\eta+2\omega\sin\eta)\left[(4+\omega^2)\cos\eta-3\omega\sin\eta+2(1+\omega^2)\right]-$$
$$\qquad\qquad-6\left[(\omega^2-1)\cos\eta+2\omega\sin\eta-\sin^2\eta+\omega^2\cos\eta\right]=$$
$$=-2(1+\omega^2)(1+7\cos\eta+2\omega\sin\eta+7\cos^2\eta+\omega\sin\eta\cos\eta),$$
and
$$J_2= -\left|\begin{array}{cc}-3 & \sin\eta-\omega\cos\eta\\
\omega+2(\sin\eta+\omega\cos\eta) & \cos\eta+\omega\sin\eta\end{array}\right|-$$
$$\qquad\qquad
-2(2+\cos\eta-\omega\sin\eta)\left|\begin{array}{cc}
2+\cos\eta+\omega\sin\eta & \sin\eta-\omega\cos\eta\\
\omega-(\sin\eta-\omega\cos\eta) & \cos\eta+\omega\sin\eta\end{array}\right|=
$$
$$=-\left[(\omega^2-3)\cos\eta-4\omega\sin\eta-2\sin^2\eta+2\omega^2\cos^2\eta\right]-$$
$$\qquad\qquad-(4+2\cos\eta-2\omega\sin\eta)\left[(2+\omega^2)\cos\eta+\omega\sin\eta+1+\omega^2\right]=
$$
$$=-(1+\omega^2)(2+7\cos\eta-2\omega\sin\eta+6\cos^2\eta-2\omega\sin\eta\cos\eta).$$
Thus the condition $\det A=0$ is written as
\begin{equation}\label{det}
\!\!\!\!\!\!\!\!\!\!\!\!
2\rho(1+7\cos\eta+2\omega\sin\eta+ 7\cos^2\eta +\omega\sin\eta\cos\eta)+
r(2+7\cos\eta-2\omega\sin\eta+ 6\cos^2\eta-2\omega\sin\eta\cos\eta)=0.
\end{equation}

Under condition $tr A=0$, matrix $A$ can be rewritten as
$$\left(\begin{array}{ccc}
-r & -2r(2+\cos\eta-\omega\sin\eta) & 2 r\rho (\sin\eta+\omega\cos\eta)\\
-\rho (2+\cos\eta+\omega\sin\eta) & -3\rho & r\rho(\sin\eta-\omega\cos\eta)\\
-\omega+\sin\eta-\omega\cos\eta & \omega+2(\sin\eta+\omega\cos\eta) & r+3\rho
\end{array}\right).$$
Its main second-order minors are
$$\left|\begin{array}{cc}
-r & -2r(2+\cos\eta-\omega\sin\eta) \\
-\rho (2+\cos\eta+\omega\sin\eta) & -3\rho
\end{array}\right|=r\rho(2\omega^2\sin^2\eta -5-8\cos\eta-2\cos^2\eta),$$
$$\left|\begin{array}{cc}
-r &  2 r\rho (\sin\eta+\omega\cos\eta)\\
-\omega+\sin\eta-\omega\cos\eta &  r+3\rho\end{array}\right|=-r(r+3\rho)-2r\rho(\sin^2\eta-\omega^2\cos^2\eta-\omega\sin\eta
-\omega^2\cos\eta),$$
and
$$\left|\begin{array}{cc}
 -3\rho & r\rho(\sin\eta-\omega\cos\eta)\\
 \omega+2(\sin\eta+\omega\cos\eta) & r+3\rho\end{array}\right|=-3\rho(r+3\rho)-r\rho(2\sin^2\eta-2\omega^2\cos^2\eta+\omega\sin\eta
 -\omega^2\cos\eta).$$
Thus, the sum $\sigma A$ of these minors vanishes (under condition $tr A=0$) when
\begin{equation}\label{sigm}
(r+3\rho)^2+r\rho(9-2\omega^2+(8-3\omega^2)\cos\eta -\omega\sin\eta -2(1+\omega^2)\cos^2\eta)=0.
\end{equation}

We are looking for values of $r,\rho,\eta,\omega$ which solve the system of Eqs.
(\ref{equ}),(\ref{tra}),(\ref{det}),(\ref{sigm}). From Eq. (\ref{tra}) we find
\begin{equation}\label{eq5}
\rho = r \frac{-1 + \cos\eta + \omega \sin\eta}{3 - 2 \cos\eta+2\omega\sin\eta}.
\end{equation}

By plugging this into Eqs. (\ref{det}) and (\ref{sigm}), we obtain the following system of equations
for $q=\omega\sin\eta$ and $z=\cos\eta$ (so $\omega^2=\frac{q^2}{1-z^2}$):
$$
2(-1 + z + q)(1+7z+7z^2+2q+qz)+(3-2z+2q)(2+7z+6z^2-2q-2qz)=0,
$$
and
$$
(z+5q)^2+(-1 + z + q)(3 - 2z+2q)(9-2\frac{q^2}{1-z^2}+(8-3\frac{q^2}{1-z^2})z -q -2(1+\frac{q^2}{1-z^2})z^2)=0.
$$
This recasts as
\begin{equation}\label{ysq}
2q^2z-q(32z^2+28z-4)-2z^3-4z^2-5z-4=0,
\end{equation}
and
$$(1-z^2)(z^2+10yz+25q^2)+(2q^2+q - 2z^2+5z-3)(9+8z-11z^2 -8z^3+2z^4 -q(1-z^2) -(2+3z+2z^2)q^2)=0,$$
or
$$2(2z^2+3z+2)q^4+q^3(4+3z)-q^2(8z^4-20z^3-51z^2+15z+48)+$$
$$\qquad\qquad +q(z^2-1)(12+13z)+(z^2-1)(4z^4-26z^3+29z^2+21z-27)=0,$$
or
$$\!\!\!\!\!\!\!\!\!\!\!\!\!\!\!\!(2z^2(2z^2+3z+2)q^2+z(64z^4+152z^3+143z^3+48z-8)q+1020z^6+3362z^5+4365z^4+2482z^3+232z^2-200z+16)\times$$
$$\qquad\times (2q^2z-q(32z^2+28z-4)-2z^3-4z^2-5z-4)=$$
$$\qquad\qquad=(32768z^8+136704z^7+230976z^6+189904z^5+60933z^4-9076z^3-5864z^2+1216z-64)q+$$
$$\qquad\qquad\qquad+2048z^9+10752z^8+27328z^7+43408z^6+45553z^5+30356z^4+10374z^3-8z^2-720z+64.$$
By Eq. (\ref{ysq}), this gives us
\begin{equation}\label{oms}
\!\!\!\!\!\!\!\!\!\!\!\!\!\!\!\!\omega\sin\eta = q=-\frac{2048z^9+10752z^8+27328z^7+43408z^6+45553z^5+30356z^4+10374z^3-8z^2-720z+64}
{32768z^8+136704z^7+230976z^6+189904z^5+60933z^4-9076z^3-5864z^2+1216z-64}.
\end{equation}
By inserting this expression into Eq. (\ref{ysq}) we obtain the following equation for $z$:
$$2(2048z^9+10752z^8+27328z^7+43408z^6+45553z^5+30356z^4+10374z^3-8z^2-720z+64)^2z+$$
$$+(2048z^9+10752z^8+27328z^7+43408z^6+45553z^5+30356z^4+10374z^3-8z^2-720z+64)\times$$
$$\qquad\qquad\qquad\times(32768z^8+136704z^7+230976z^6+189904z^5+60933z^4-9076z^3-5864z^2+1216z-64)\times$$
$$\qquad\qquad\qquad\qquad\qquad\times(32z^2+28z-4)-$$
$$-(2z^3+4z^2+5z+4)(32768z^8+136704z^7+230976z^6+189904z^5+60933z^4-9076z^3-5864z^2+1216z-64)^2=0,$$
or
$$P(z)=
-512-1536z+123584z^2-294656z^3-2669168z^4-1014068z^5+16713471z^6+38944576z^7+$$
$$\quad+29896640z^8-11432448z^9
-38759936z^{10}-28958720z^{11}-8060928z^{12}+524288z^{13}+524288z^{14}=0.$$
This polynomial has the following roots in the interval $[-1,1]$:
$$z\approx -.8468602601, -.8453251846, -.05672395050, .09599192317, .1369140710, .2980830761, .982719862$$
The root $z=\cos\eta=z_0\approx .136914071$ corresponds to
$$\eta=\eta_0\approx 1.433450854.$$
By Eq. (\ref{oms}) one finds
$$\omega=\omega_0\approx -3.487162,$$
and, by Eqs. (\ref{equ}),(\ref{eq5}),
$$r=r_0\approx 0.092903423, \quad \rho=\rho_0\approx 0.09590066.$$
One may check that the other roots of $P(z)$ do not produce positive values of $r$ and $\rho$.
The corresponding values of $\beta=\beta_0$ and $\gamma=\gamma_0$ are found from Eq. (\ref{bgm}):
$$\beta_0\approx 1.16531, \gamma_0\approx 0.224354$$

Note that $P'(z_0)\approx -6633\neq 0$, therefore any small perturbation
of the system of Eqs. (\ref{tra}),(\ref{det}),(\ref{sigm}),(\ref{equ})
will have a solution $(r,\rho,\omega,\eta)$ close to $(r_0,\rho_0,\eta_0)$. Thus, for
any given small values $\mu,\nu,\lambda$, we can always find
values of parameters $(\beta,\gamma,\omega)$ close to $(\beta_0,\gamma_0,\omega_0)$ which would correspond to
the existence of an equilibrium state close to $(r_0,\rho_0,\eta_0)$ such that the linearization matrix $A$
at this equilibrium will have $tr A=\mu$, $\sigma A=\nu$ and $det A=\lambda$.\\

Consider system (\ref{ssm}) at $(\beta,\gamma,\omega)=(\beta_0,\gamma_0,\omega_0)$. We put the coordinate origin to the equilibrium,
i.e. we denote $x_1=r-r_0,x_2=\rho-\rho_0,x_3=\eta-\eta_0$. The system takes the form
$$\dot x=Ax+F(x)+O(\|x\|^3)$$
where $F(x)$ contains only quadratic terms; recall that the matrix $A$ has three zero eigenvalues, so $A^3=0$. We take
the vectors $A^2e$, $Ae$ and $e=\left(\begin{array}{c} 0\\ 1\\ 0\end{array}\right)$ as the coordinate basis.
In other words, we make a coordinate transformation $x=QX$, where $Q=(A^2e,Ae,e)$. One needs to check that $det(Q)\neq 0$;
then the system takes the form
$$\frac{d}{dt}\left(\begin{array}{c} X_1\\ X_2 \\ X_3\end{array}\right)=
\left(\begin{array}{c} X_2\\ X_3 \\ 0\end{array}\right)+ Q^{-1}F(QX)+O(\|X\|^3).$$
Moreover, when we add any small perturbation to this system we can always choose coordinates so that
$\dot X_1=X_2$ and $\dot X_2=X_3$. Thus, if a perturbation of the $C^2$-size $\delta$ is added so that the equilibrium state
does not disappear, the system takes the form
\begin{equation}\label{ACTS}
\frac{d}{dt}\left(\begin{array}{c} X_1\\ X_2 \\ X_3\end{array}\right)=
\left(\begin{array}{c} X_2\\ X_3 \\ \lambda X_1 -\nu X_2+\mu X_3\end{array}\right)+ Q^{-1}F(QX)+O(\|X\|^3+\delta\|X\|^2).
\end{equation}
As we just mentioned, the coefficients $\mu =tr A$, $\nu=\sigma A$ and $\lambda=det A$ can acquire arbitrary small values
when the parameters $\beta,\gamma,\omega$ are changed appropriately. It has been shown in Refs. \cite{Tresser,Ibanez} that systems of
form (\ref{ACTS}) have chaotic dynamics (the Shilnikov saddle-focus loop) at some values of $\mu,\nu,\lambda$ (which can be chosen
arbitrarily small), provided the the coefficient $a$ of $X_1^2$ in the equation for $\dot X_3$ is not zero.

Thus, to prove the existence of chaotic behavior in system (\ref{ssm}), it remains to compute $det Q$ and $a$.
By plugging the values of $r_0,\rho_0,\eta_0,\omega_0$ into Eq. (\ref{matr}), we find
$$A\approx\left(\begin{array}{ccc}
-0.09290342 & -1.0388902 & 0.009143666\\ 0.1263404 & -0.287702 & 0.01307936\\
4.955187 & -2.460879 & 0.3806054 \end{array}\right),$$
$$A^2\approx\left(\begin{array}{ccc}
-0.07731418 & 0.3729058 & -0.01095737\\
0.01672483 & -0.080668085 & 0.0023703315\\
1.1147085 & -5.376519 & 0.1579823\end{array}\right).$$
Thus, the matrix $Q=(A^2e,Ae,e)$ is given by
$$Q\approx\left(\begin{array}{ccc}
0.3729058 & -1.0388902 & 0\\
-0.080668085 & -0.287702 & 1\\
-5.376519 & -2.460879 & 0\end{array}\right).$$
We have $det Q\approx 6.503289\neq 0$, so the system can indeed be brought to form (\ref{ACTS}).

It remains to find the coefficient $a$ of $X_1^2$ in the equation for $\dot X_3$ in Eq. (\ref{ACTS})). It equals to
the product of the third row of the matrix $Q^{-1}$ to $f(A^2e)$. The third row of $Q^{-1}$ is orthogonal to the
first and second columns of $Q$, i.e. to the vectors $A^2e$ and $Ae$. Any row of the matrix $A^2$ satisfies this property
(recall that $A^3=0$). Therefore, in order to check that $a\neq 0$, it is enough to check that the product $f(A^2e)$ to the third
row of $A^2$ is non-zero. Since $f$ is the quadratic part of the Taylor expansion of the right-hand side of system (\ref{ssm}) at
the point $(r_0,\rho_0,eta_0)$, we need to compute the coefficient of $\varepsilon^2$ in the expansion in powers of $\varepsilon$
for $p_1\dot r+p_2\dot \rho+p_3\dot\eta$ evaluated at $(r,\rho,\eta)=(r_0+q_1\varepsilon,\rho_0+q_2\varepsilon,\eta_0+q_3\varepsilon)$,
where $(p_1,p_2,p_3)$ is the third row of $A^2$ and $(q_1,q_2,q_3)^\top=A^2e$ is the second column of $A^2$;
recall that
$$p_1\approx 1.1147085, \qquad p_2 \approx -5.376519 , \qquad p_3\approx  0.1579823,$$
$$q_1\approx 0.3729058, \qquad q_2 \approx -0.080668085, \qquad q_3\approx -5.376519.$$
We have
$$p_1\dot r+p_2\dot \rho+p_3\dot\eta= -p_1 r^2-3p_2\rho^2-r\rho(4p_1+2p_1(\cos\eta-\omega\sin\eta)+p_2(2+\cos\eta+\omega\sin\eta))+$$
$$+p_3(\sin\eta-\omega\cos\eta)r+2p_3(\sin\eta+\omega\cos\eta)\rho+\mbox{ linear terms}.$$
By plugging $(r,\rho,\eta)=(r_0+q_1\varepsilon,\rho_0+q_2\varepsilon,\eta_0+q_3\varepsilon)$ in the right-hand side,
we find that the coefficient of $\varepsilon^2$ equals to
$$-p_1q_1^2-3p_2q_2^2-q_1q_2(4p_1+2p_2+(2p_1+p_2)\cos\eta_0+(p_2-2p_1)\omega_0\sin\eta_0)+$$
$$+(q_1\rho_0+q_2r_0)q_3((2p_1+p_2)\sin\eta_0+(2p_1-p_2)\omega_0\cos\eta_0)+\frac{1}{2}r_0\rho_0q_3^2((2p_1+p_2)\cos\eta_0
+(p_2-2p_1)\omega_0\sin\eta_0)-$$
$$-\frac{1}{2}p_3q_3^2(\sin\eta_0-\omega_0\cos\eta_0)r_0+p_3q_3q_1(\cos\eta_0+\omega_0\sin\eta_0)
-p_3q_3^2(\sin\eta_0+\omega_0\cos\eta_0)\rho_0+$$
$$+2p_3q_3q_2(\cos\eta_0-\omega_0\sin\eta_0)\approx 5.898,$$
i.e. it is non-zero. This finishes the proof of the existence of chaotic dynamics in system (\ref{ssm}).

\end{document}